\documentclass[12pt,manuscript]{aastex}

\def\ngc#1{\hbox{NGC$\,$#1}}

\def\etal{{et~al.}}
\def\ie{{\it i.e.}}
\def\eg{{\it e.g.}}

\def\rnum#1{{\uppercase\expandafter{\romannumeral#1}}}

\def\mag{$\,$mag}

\def\hr{${}^{\rm h}$}
\def\mn{${}^{\rm m}$}

\def\Sc{${}^{\rm s}$\llap{.}}

\def\Min{${}^{\prime}$\llap{.}}
\def\Sec{${}^{\prime\prime}$\llap{.}}
\def\deg{${}^\circ$}
\def\min{${}^{\prime}$}

\def\gtsim{ \,{}^>_\sim\, }

\def\vmi{\hbox{\it V--I\/}}
\def\vmr{\hbox{\it V--R\/}}
\def\rmi{\hbox{\it R--I\/}}
\def\bmv{\hbox{\it B--V\/}}
\def\umb{\hbox{\it U--B\/}}
\def\bmi{\hbox{\it B--I\/}}

\def\today{\number\year\space \ifcase\month\or
  January\or February\or March\or April\or May\or June\or
  July\or August\or September\or October\or November\or December\fi
  \space\number\day}
\def\now{\number\year\space \ifcase\month\or
  January\or February\or March\or April\or May\or June\or
  July\or August\or September\or October\or November\or December\fi
  \space\number\day .\number\time}
\begin{document}

\received{}
\revised{}
\accepted{}
\ccc{}
\cpright{}{}

\slugcomment{}
\shorttitle{\ngc{188}}
\shortauthors{Stetson, McClure, VandenBerg}

\title{A Star Catalog for the Open Cluster \ngc{188}}

\author{Peter B. Stetson\altaffilmark{1,}\altaffilmark{2} and Robert D. McClure}
\affil{Dominion Astrophysical Observatory, Herzberg Institute of Astrophysics,
National Research Council, 5071 West Saanich Road, Victoria, BC V9E 2E7, Canada;
Peter.Stetson@nrc.gc.ca}

\author{and}

\author{Don A. VandenBerg}
\affil{Department of Physics and Astronomy, University of Victoria, Box 3055,
Victoria, BC V8W 3P6, Canada; davb@uvvm.uvic.ca}

\altaffiltext{1}{Guest Investigator of the UK Astronomy Data Centre.}

\altaffiltext{2}{Guest User, Canadian Astronomy Data Centre, which is operated by the
Herzberg Institute of Astrophysics, National Research Council of Canada.}

\begin{abstract}

We present new {\it BVRI\/} broad-band photometry for the old open cluster
\ngc{188}\ based upon analysis of 299 CCD
images either obtained by us, donated by colleagues, or retrieved from public
archives.  We compare our results on a star-by-star basis with data from eleven
previous photometric datasets for the cluster.  We homogenize and merge the data
from all the photometric studies, and also merge membership probabilities from
four previous proper-motion studies of the cluster field.  Fiducial cluster
sequences in the $BV$ (Johnson) $RI$ (Cousins) photometric system of Landolt
(1992, \aj, 104, 340) represent the principal result of this paper.  These have
been compared to reference samples defined by (a)~Landolt's standard stars,
(b)~the old open clusters M$\,$67 and \ngc{6791}, and (c)~stars within 25$\,$pc
having modern photometry and precise {\it Hipparcos\/} parallaxes.  In a
companion paper we show that our derived cluster results agree well with the
predictions of modern stellar-interior and -evolution theory, given reasonable
estimates of the cluster chemical abundances and foreground reddening.  
The individual and combined datasets for \ngc{188}\ have been made available
through our web site.  

\end{abstract}

\keywords{Open clusters and associations: individual}

\section{INTRODUCTION}

The study of star clusters provides important clues to the formation and
enrichment history of the Milky Way galaxy.  To the extent that we can consider
the different stars in a given cluster to have a common distance, age, chemical
content, and foreground extinction they provide stronger constraints on models
of stellar evolution than field stars, whose distances, reddenings, and
metallicities must be estimated on an individual basis.  While studies of field
stars can provide important insights into extremes of the local stellar
population (the oldest, the most metal-rich, the most metal-poor stars; \eg,
Sandage, Lubin \& VandenBerg 2003), in the middle of the Hertzsprung-Russell
diagram there is such a congestion of stars with a range of ages,
metallicities, and evolutionary states that extremely high accuracy {\it and\/}
precision in estimates of effective temperature, metal abundance, and distance
are required to derive a reliable age estimate for any given isolated star.

The study of star clusters relaxes the need for extreme accuracy and precision,
at least to some extent.  Spectroscopic abundance determinations can be derived
for a number of different stars in the same cluster, and the average of those
determinations can be applied to the cluster as a whole with a precision that is
probably higher than possible for any single star.  More important, the presence
within a cluster of stars of different masses allows us to estimate a common
cluster age and distance by forcing consistency with a given set of theoretical
isochrones over a range of luminosities and temperatures, and hence of
evolutionary states.  At least to the extent that relative ages among different
clusters can be estimated from the {\it shapes\/} of ther principal squences
(\eg, Anthony-Twarog \& Twarog 1985; VandenBerg, Bolte \& Stetson 1990;
Sarajedini \& Demarque 1990; Stetson, Bruntt \& Grundahl 2003), uncertainties in
distance and reddening become less critical.  

Ever since it was first noted that the open cluster \ngc{188} was likely to be
very old (van~den~Bergh 1958 appears to be the first published mention, although
Sandage 1962 has noted that Ivan King apparently recognized the particular
interest of this cluster in 1948), it has been the subject of numerous
photometric and astrometric studies.  Photometric observations that have made
significant contributions to our understanding of \ngc{188} were reported by
Sandage, Sharov (1965), Cannon (1968), Eggen \& Sandage (1969), McClure \&
Twarog (1977), Caputo \etal\ (1990), Dinescu \etal\ (1996), von~Hippel \&
Sarajedini (1998), Sarajedini \etal\ (1999), and Platais \etal\ (2003).  In
addition, Kaluzny \& Shara (1987); Zhang \etal\ (2002); and Kafka
\& Honeycutt (2003) have published photometric studies of variable stars in the
cluster field.  Finally, Cannon; Upgren, Mesrobian \& Kerridge (1972); Dinescu
\etal; and Platais \etal\ have published membership probabilities for stars in
the \ngc{188} field based upon measurements of their proper motions.

In the present paper we present newly derived photometric indices for stars
in \ngc{188} from CCD images which we have either obtained ourselves,
received from colleagues, or recovered from public-domain archives.  These
photometric results are compared and then combined on a star-by-star basis with
photometric data from previous studies of the cluster.  We also merge the
results of previous astrometric studies of the cluster and combine them 
with the photometric information to provide color-magnitude and color-color
diagrams of probable cluster members.\footnote{The present paper and a
companion paper by VandenBerg \& Stetson represent a fulfillment and extension
of the paper ``D.~A.~VandenBerg \& R.~D.~McClure (2003, in preparation)'' cited
by Sandage \etal\ (2003) in their critical comparison of the old open clusters
M$\,$67, \ngc{188}, and \ngc{6791} to the old field stars of the Solar
Neighborhood.} 

\section{OLD OBSERVATIONS AND NEW}

\subsection{Photometry}

As mentioned in the Introduction, \ngc{188} has been the subject of numerous
previous photometric studies.  Table~1 presents an in-a-glance summary of the
properties of the previous and present photometric studies of the cluster.  Here
and henceforth in this paper, when discussing a particular photometric dataset
(as opposed to a paper) we will save space by labeling that dataset with the
name of the first author of the publication only, with terse modifiers when
needed.  For each dataset which we consider distinct (column 1 of the table), we
list the technology used to acquire the data (column 2), the number of stars for
which photometric data are presented (column 3), the limiting magnitudes in $U$,
$B$, $V$, and $I$ (columns 4 through 7), and the source of the data which were
employed to calibrate the photometry of the given study (column 8).  Here, we
define ``limiting magnitude'' simply as the 95${}^{\hbox{\footnotesize th}}$
percentile of the magnitudes listed in the dataset; this is intended purely as a
crude indicator of the greatest depth effectively probed, and we make no claims
concerning the completeness of any sample.  We have not listed the limiting
magnitude in $R$, but each of the two studies that provides $I$ magnitudes also
gives $R$ magnitudes to a corresponding magnitude limit.  The Upgren study is
included in this table for completeness, but most of the photometry presented in
that paper was not original with them; rather they tabulated a compilation of
previous values taken from the literature, with magnitudes estimated from their
plates only when data from other sources were not available.  We were unable to
locate copies of the original von~Hippel \& Sarajedini data tables.  This study
is therefore not included in the analysis which follows.  

We obtained 86 CCD images of \ngc{188} with the f/5 Newtonian camera on the
Plaskett 1.8m telescope of the Dominion Astrophysical Observatory on the nights
of 1996 July 11 and 12 (UT).  The CCD camera covered a field 9\Min3 square
with a sampling of 0\Sec55 per pixel.  We took 28 images in the $B$ photometric
bandpass, 33 in $V$, and 25 in $R$ divided among three separate subfields, and
exposure times ranged from 6$\,$s to 900$\,$s.  Conditions were judged to be
non-photometric when these observations were made.  During the same nights
observations were obtained for eight different equatorial standard fields from
Landolt (1992).  In addition, we acquired from the DAO archive copies of images
obtained by Manuela~Zoccali using the same equipment on the nights of July 15
and 16; she had observed secondary standard fields in the globular cluster M92
and the old open cluster \ngc{6791} (see Stetson 2000 and Stetson \etal\ 2003). 
Because all these data were obtained under non-photometric conditions, we are
not able to fully calibrate them to a fundamental magnitude system on the basis
of internal information alone.  However, we can exploit the fact that stars of
differing colors were imaged on the detector simultaneously to determine the
color-dependent terms in the transformation equations from the instrumental to
the standard magnitude system.  Later on we will use the results of previous
photometric studies of the cluster to determine the calibration zero-point for
each of the individual images.

Howard~Bond has given us copies of his images from 13 different observing
runs at the Kitt Peak 4m and 0.9m and the Cerro Tololo 0.9m telescopes.  
Among the Kitt Peak 0.9m data were 12 images of NGC188---four in each of the
$B$, $V$, and $I$ filters---obtained on 1996 September 21.  We also
searched the available public archives for other images of \ngc{188} and found
201 of them: 6 exposures with the four-chip Isaac Newton Telescope Wide-Field
Camera (24 images in total), and a further 177 images obtained with single-CCD
cameras from three different observing runs on the INT.  Of these data, only two
observing runs were judged to be photometric on the basis of standard-star
calibration residuals, a point which we will discuss in greater detail below.  

We requested from the ING archive all images---both scientific and
calibration---from those observing runs when \ngc{188} was observed,
and derived and applied the bias and flat-field corrections in the usual
way.  This was also done for the Plaskett data, of course, but Howard
Bond had provided us his images in already-rectified form.  The DAOPHOT,
ALLSTAR, DAOMATCH, DAOMASTER, and ALLFRAME software packages (Stetson 1987,
1993, 1994) were used to detect and measure the positions and brightnesses of
astronomical sources in all 299 of the CCD images that were available to us. 
The star-subtracted images were stacked and examined by eye to designate stars
which had been missed by the automatic routines, and the ALLFRAME reductions
were repeated until we judged that everything that we could see had been
reduced.  

The images were then subjected to a routine growth-curve analysis to determine
corrections from the profile-fitting magnitudes to a system of aperture
magnitudes in a large synthetic aperture (Stetson 1990), and calibration
equations relating the instrumental-system magnitudes to the standard system of
Landolt were derived.  Full transformations, complete with zero-points and
extinction coefficients as well as linear and quadratic color terms, were
possible only for the two INT observing runs where apparently photometric
conditions prevailed.  For the remaining data it was possible to determine only
the color terms of the transformation equations.

\subsection{Astrometry}

In order to compile a master star list for the region of \ngc{188}, we
supplemented our own CCD images with ten other star lists from the archives and
literature.  (1) We extracted from the U.~S.~Naval Observatory ``USNO-A2.0''
Guide-Star Catalog (Monet \etal\ 1998) all 9,503 sources within a square box
66\Min0 on a side centered on coordinates $\alpha\,=\,00$\hr47\mn27\Sc52,
$\delta\,=\,+85$\deg16\min10\Sec7 (J2000).  (2)--(5) Employing the services
of the Canadian Astronomy Data Centre, we extracted images 54\min\
on a side, centered on the same coordinates, from the Digital Sky Survey~1 ``O''
plate, and the Digital Sky Survey~2 ``B,'' ``R,'', and ``I'' plates.  These were
analyzed with a modernized version of the Stetson (1979) software.  (6)~Cannon,
(7)~Upgren \etal, (8)~Dinescu \etal\, (9)~Sarajedini \etal\ (private
communication), and (10)~Platais \etal\ all provided positional determinations
along with their photometric indices.  The program DAOMASTER was then used to
transform the data from these star lists and our own ALLFRAME analysis of the
299 CCD images to a common reference system and numbering scheme.  Ten-parameter
cubic fits in $x$ and $y$ were used to effect the transformations.  The USNO2
positions were used as the primary reference, so the $(x,y)$ coordinates in our
composite star list should be accurately aligned with the cardinal directions,
with $x$ increasing east and $y$ increasing north.  Positions are expressed in
units of arcseconds with the origin of the coordinate system at the celestial
coordinates given above.  A critical match-up tolerance was gradually decreased
from 9 to 4 arcseconds in deciding whether entries in different star lists
referred to the same object.  This tolerance was larger than we normally employ,
but in this case it was necessary in order to deal with the Upgren \etal\
positions, which were given only to the nearest one-tenth arcminute.  Table~2
indicates the number of stars from each sample which were successfully merged
into the master list, and the precision of the positions as inferred from the
fitting residuals.  It would be unfair to interpret these precision estimates as
reflecting the relative astrometric accuracy or precision of the various
studies, since different researchers probably adopted different tolerances for
the degree of crowding or the limiting signal-to-noise ratio for the detections
they chose to tabulate, and the $\sigma$'s here will be strongly influenced by
the {\it worst\/} data each study chose to report.  The point here is that when
appropriate weighted combinations are formed from the reported positions of 
objects in the NGC188 field, we can generally expect them to be precise to well
under 0\Sec1.

\section{(RE)CALIBRATION AND COMBINATION}

Calibrating photometry for \ngc{188} is unusually difficult.  The cluster lies
at a declination of +85\deg, which means that it stays at a high airmass
($\gtsim\,$1.8) from sites in the southern United States and the Canary Islands,
regardless of the time of night or time of year.  This declination also places
the cluster just about as far as it can possibly be from the best-established
faint photometric standard stars, which are concentrated near the equator
and in the southern hemisphere.

We have at hand seven datasets that were obtained under nominally photometric
conditions, so they can be used to estimate the zero-point of the magnitude
scale in the field of NGC188:  (1)~the photoelectric {\it UBV\/} photometry of
Sandage (1962); (2)~the photoelectric photometry attributed to Eggen in Eggen \&
Sandage (1969); (3)~the photoelectric photometry attributed to Sandage in the
same paper; (4)~the CCD photometry of Sarajedini \etal\ (1999), who had one
photometric night during which they referred their \ngc{188} data to 38 of
Landolt's (1992)~equatorial standards; (5)~the CCD photometry of Platais \etal\
(2003) which, they state, was calibrated by unpublished photometric observations
attributed to Hainline \etal\ (2000) and employed color terms privately provided
by P.~Massey---we can only presume that these were based upon Landolt and/or 
Johnson standards as are other WIYN Open Cluster Studies; and (6)~and (7), two
of the datasets which we obtained through the services of the Isaac Newton Group
Archive, specifically ones which we designate cmr2 and wkg (labels based upon
the initials of the observer, which are the only personal identifiers available
from the image headers).  Of these, the former comprised a single night of
photometric observations (1993 June 24/25) from which we were able to obtain
5,673 individual measurements in $V$ and 4,678 in $B$ of primary and secondary
standard stars from Stetson's on-line compilation (see Stetson 2000), plus three
images in each of $B$ and $V$ for a single 12\Min5$\times$12\Min5 field in
NGC188.  The latter consisted of three nights of observations (1987 Jul 31/Aug 1
to Aug 2/3) providing from 550 to 1,100 standard-star measurements in each
filter on each night, plus 139 images of nine distinct 6\Min3$\times$3\Min9
(east-west $\times$ north-south) sub-fields in the cluster.  Any given sub-field
had as few as two observations in each of $B$ and $V$, or as many as eight
observations in each of $B$, $V$, $R$, and $I$.  (Interested readers can recover
the placement of these fields on the sky from the stellar positions and
cross-identifications that we have posted at our web site, or by querying the
ING archive\footnote{ http://archive.ast.cam.ac.uk/ingarch/ingarch.html} for
observations obtained with the INT on the dates indicated above.)  For each of
these ING datasets, {\it ex post facto\/} consideration of the standard-star
residuals from the photometric calibration solution suggests that conditions
were photometric at a level of 0.01--0.03$\,$mag per observation.  Accordingly,
zero-points derived from the cmr2 data alone, for instance, are not likely to be
more accurate than $\sim\,{0.02\,\hbox{\footnotesize mag}\over\sqrt{3}}
\sim\,$0.01--0.02\mag, since in the final analysis they are based on only three
observations in each filter, and the photometric errors associated with those
observations are not likely to be completely independent. 

It is possible that the Hainline \etal\ observations alluded to by Platais 
\etal\ are actually the same as those reported by Sarajedini \etal: both
were obtained with the Kitt Peak 0.9m telescope, and some but not all
authors are common to the published abstract and the two refereed papers. 
However, Platais \etal\ state that the Hainline \etal\ observations span
40~arcminutes on the sky, whereas the Sarajedini \etal\ data span only about
23~arcminutes.  Furthermore, as we will show below, comparison of the Platais
and Sarajedini magnitudes shows both a small net offset and considerable scatter
in the differences.  We therefore assume the two datasets to be statistically
independent.  

It is also not immediately obvious that datasets (1), (2), and (3) are mutually
independent.  In describing the origins of the data in their Table~1, Eggen \&
Sandage (1969) state that they have tried to ``make the best combination of new
200-inch observations by Eggen with the older data by Sandage'' by determining
and removing systematic differences between the two datasets.  The ``older data
by Sandage'' would most obviously mean the data from Sandage (1962).  However,
there is no obvious one-to-one correspondence between the data from the earlier
paper and the observations attributed to Sandage in the later one:  Sandage
(1962) reports photoelectric results for 123 stars, while the observations
attributed to Sandage in Eggen \& Sandage (1969) represent only 97.  There are
28 stars that are in the first paper but not the second, and two stars (both
named ``Anon'') in the second paper but not the first.   A comparison of the
magnitude values for stars in common to the two datasets indicates net offsets
of $\sim\,$0.03\mag\ in the {\it UBV\/} magnitude systems which could, of
course, be the result of the recalibration undertaken in the 1969 paper. 
However, the star-by-star comparison also shows dispersions in the magnitude
differences of $\sim\,$0.05\mag\ in $U$ (72 stars in common) and
$\sim\,$0.04\mag\ in $B$ and $V$ (95 stars) between the Sandage (1962)
magnitudes and the ``S'' magnitudes of Eggen \& Sandage (1969).  It is evident
that the latter are not simply the former with an additive calibration
correction applied.  Furthermore, plots of $\Delta V$ and $\Delta B$ (Sandage
1962 {\it versus\/} Eggen \& Sandage 1969 ``S'') against \bmv\ show no clear
trend, so we are not dealing with a simultaneous adjustment of zero-point and
color transformation.  Changes to the extinction correction cannot explain the
dispersion in the magnitude differences, since \ngc{188} lies at a virtually
constant airmass, so changes to the extinction coefficients should affect all
stars in the field equally. Instead, the two datasets {\it seem\/} to derive
from different observations.  
Similarly, the Eggen \& Sandage ``E'' stars that are in common with the ``S''
stars in the same table show a net offset of 0.02$\,$mag in $V$ and 0.01$\,$mag
in each of $B$ and $U$ (17 stars in {\it BV\/} and 16 in $U$), so it is not
obvious that the latter have been placed on the magnitude system of the former
by a simple zero-point adjustment.  Therefore, for the purposes of this paper we
will assume that these three datasets, Sandage (1962) and the Eggen \& Sandage
(1969) ``E'' and ``S'' data, are all mutually independent.  

Under this assumption, then, we have seven mutually independent attempts to
establish a fundamental magnitude scale in the field of NGC188.  The three
oldest datasets (Sandage, Eggen ``E,'' and Eggen ``S'') were explicitly
calibrated to the {\it UBV\/} system of Johnson \&\ Morgan (1953) and Johnson
(1955); the fourth and fifth (Sarajedini and Platais) employed the standard
photometric system of Landolt (1992), who did his best to place his {\it UBV\/}
magnitudes on the system of Johnson \& Morgan and Johnson (see Landolt 1983);
and for the last two photometric datasets (cmr2 and wkg) we employed the
standard stars tabulated by Stetson\footnote{
http://cadcwww.hia.nrc.ca/cadcbin/wdb/astrocat/stetson/query/}
(see Stetson 2000), who is doing his best to ensure that his magnitudes are on
the same system as Landolt's.  Therefore, we will further assume that the seven
datasets represent independent attempts to establish the {\it same\/} magnitude
scale, and that they differ only as a result of systematic errors which vary at
random among the different datasets.  

To determine the net study-to-study systematic magnitude offsets in order to
define an optimum average system, we began by using the Sarajedini \etal\
photometry as an initial reference, since that dataset is unique in including
measurements in all five of the {\it UBVRI\/} bandpasses.  The top panel of
Table~3 lists the unweighted arithmetic mean  magnitude differences between
each of the other six studies and the Sarajedini \etal\ results in those
filters for which data are available.  (The Platais study lists the $V$
magnitude of star III-138 as 12.07.  For this star, the Sandage (photoelectric),
Eggen ``S'', McClure, and our own datasets all give $V$ in the range 9.71 to
9.80.  The Platais data for this star are not used here.)  Only the Platais
data, the Sarajedini data, and our rereductions of the ING archival data
provide a standard-error estimate for each derived standard-system magnitude; we
have used these to reject any measurement with a claimed uncertainty
$\geq\,$0.10\mag, but apart from this no use was made of the standard errors in
weighting the magnitude differences.  The three photoelectric datasets from the
1960's provided no individual error estimates, and all published measurements
were retained in the analysis except that it was necessary to manually exclude
variable stars V5 = I-11 and V11 = I-116 from these comparisons.  We note from
the data in Table~3 that the Sarajedini \etal\ magnitudes are fainter than all
six of the other studies in both $B$ and $V$; the Sarajedini magnitudes are also
fainter than the one other available study in $R$ and $I$.

The root-mean-square standard deviation of the magnitude differences in each
comparison is provided in the second panel of the table.  Some portion of these
dispersions is the result of random errors in the individual measurements, but
it is likely that there is also some contribution from filter mismatch. 
Stetson \etal\ (2003) provided evidence that even when measurements
of arbitrarily high precision are obtained with a given filter/detector
combination, there still tends to be an irreducible scatter of order
0.01--0.02$\,$mag when these observations are transformed to a photometric
system that is nominally the same, but which was defined with a different
filter/detector combination.  (This conclusion was based on observations in
the $B$, $V$, and $I$ bandpasses.  There is no reason to assume that it does
not apply equally well to the $R$ bandpass.  The study-to-study repeatability
of $U$ magnitudes may well be worse than this, since the $U$ bandpass is
atmosphere-defined on the short-wavelength side, and moreover it includes the
Balmer convergence and jump, which are affected nonlinearly by temperature and
are also sensitive to surface gravity.)  These study-to-study differences are
not simply correlated with color (if they were, they could be removed in the
transformation), and they can not be improved by taking better data with the
same equipment.  Thus they do not depend upon the internal precision of any
given investigation.  This justifies our refusal to weight individual
measurements on the basis of their published standard errors or the number of
independent measurements in any given study as we compare the different
photometric systems.  

The third panel of Table~3 lists the median magnitude difference found
between the Sarajedini data and each of the other photometric studies.  Since
astronomical measurements are subject to many small random uncertainties and
also sometimes to large sporadic mistakes, they often do not follow a simple
Gaussian probability distribution.  When mistakes happen, a few large deviations
can dominate a traditional statistical analysis.  Therefore, we present these
median offsets because they are more ``robust'' or ``resistant'' than the
arithmetic mean---more representative of the typical magnitude difference
without being as strongly influenced by a few extreme outliers that might be due
to the inclusion of variable stars, typographical errors, or urecognized
peculiar anomalies caused by, for instance, contrails, cosmic rays, or insects
in the apparatus.  For the same reason, the fourth panel of the table lists
$\sqrt{\pi\over2}$ times the mean absolute difference between the tabulated
magnitudes in each study and the Sarajedini data after removal of the median
offset.  This is a measure of the dispersion in the magnitude differences that
is less affected by the most extreme outliers than the more commonly used
root-mean-square statistic.  If the frequency distribution of the magnitude
differences were accurately Gaussian in form, then the expectation value of the
standard deviation $\sigma$ would be equal to $\sqrt{\pi\over 2} \approx 1.2533$
times the mean absolute deviation.  

A comparison of the mean and median magnitude differences shows that they are
practically the same, suggesting little or no skewness in the
distributions---there is no evidence that any one study tended to have large
errors that were preferentially too bright or too faint.  However, the robust
estimators of dispersion in the fourth panel of the table are systematically
smaller than the standard deviations given in the second panel, indicating that
the magnitude measurements have probability distributions that are somewhat
non-Gaussian in form, with an excess of unusually large discrepancies (positive
kurtosis).  The fifth panel of the table indicates how many stars were
used in each comparison.  

The first and third panels of Table~3 should each be considered to contain
a row of zeros for the offset of the Sarajedini dataset with respect to itself.
If we include those zeros and average together the seven offsets in each filter
(in $B$ and $V$, fewer in $U$, $R$, and $I$), then we get the average offset
of the entire ensemble of seven photometric systems relative to the Sarajedini
system and the standard deviation of those systems about the mean system.  These
numbers are contained in the rows labeled ``average'' and ``standard
deviation.'' These show that whether we judge by mean or median offset, in every
filter the Sarajedini magnitude scale is fainter than the average of all seven
by amounts ranging up to $\sim\,$0.04\mag.  Accordingly, we infer that the
optimum compromise magnitude scale in each filter is one that is brighter than
the Sarajedini system by the indicated amounts.  We also see that the standard
deviation of the offsets in the different studies and filters averages of order
0.04$\,$mag.  This indicates that any one attempt to establish a magnitude scale
in the field of \ngc{188} has typically been subject to an unknown systematic
error with a standard deviation of roughly 0.04$\,$mag.  Therefore we hope that
by thus averaging the results of seven different calibration attempts, we have
been able to establish a compromise magnitude scale that is {\it probably\/} on
the true Johnson system to within ${0.04\,\hbox{\footnotesize mag}\over\sqrt{7}}
\approx 0.015\,$mag (standard error of the mean) in each of $B$ and $V$, with
correspondingly poorer confidence in $U$, $R$, and $I$.  

Examination of Table~3 reveals at most a slight tendency for the offsets in $B$
and $V$ to be correlated.  If they were in fact correlated one-to-one, it would
be possible to consider the zero-point of the \bmv\ scale to have been
established with essentially negligible uncertainty.  But in the present case,
splitting the difference between the mean and median offsets and rounding to the
nearest 0.01$\,$mag we find that the zero-point of Sandage's \bmv\ color scale
is 0.01$\,$mag bluer than Sarajedini's, the Eggen \& Sandage ``E'' colors are
0.02$\,$mag redder, the Eggen \& Sandage ``S'' data agree with Sarajedini's,
Platais's are 0.01 redder, and the cmr2 and wkg colors are 0.07$\,$mag and
0.04$\,$mag bluer, respectively.  With the data as they are, it seems a
worse-case scenario is more likely:  while the mean of the seven color scales is
about 0.02$\,$mag bluer than Sarajedini's, $B$ and $V$ are largely uncorrelated
and the standard deviation of the different color zero-points is about
0.03$\,$mag.  Therefore we expect that the systematic uncertainty of the
zero-point of the color scale is probably $\sim\,0.01\,$mag: the absolute
Johnson \bmv\ colors of \ngc{188}'s stars will be uncertain by at least this
amount quite apart from any uncertainty in the foreground reddening.  

Splitting the difference between the mean and median offsets, we conclude that
the optimum compromise magnitudes will be brighter than Sarajedini's by 0.004,
0.041, 0.028, 0.043, and 0.028$\,$mag in $U$, $B$, $V$, $R$, and $I$,
respectively.  At the same time, for instance, the Eggen ``E''  photoelectric
$B$ magnitudes should be shifted fainter:  that is, to remove the 0.064$\,$mag
net offset between the Sarajedini $B$-band magnitudes and those of the Eggen
``E'' sample, we shift the former brighter by 0.041$\,$mag and the latter
fainter by 0.023$\,$mag, bringing both to the average system of all seven
studies.  This is done for all the datasets and filters, with the individual
corrections listed in the last section of Table~3.  Having referred every
dataset to a common magnitude system, we are now able to form unbiased averages
of the various studies' magnitudes for stars in common.  When this is done, we
find that we have a total of (896, 7519, 7819, 1549, 1543) stars with at least
one valid measurement in ($U$, $B$, $V$, $R$, $I$), respectively.  If we
restrict the sample to those stars with at least two independent measurements
in a given filter, they number (140, 1440, 1560, 299, 300).  

For our immediate purposes, we selected the 299 stars in the sample that have at
least two independent calibrated measurements in {\it each\/} of $B$, $V$, $R$,
and $I$.  These stars were then used as local in-frame standards that permitted
us to recalibrate our own CCD reductions, including images that were obtained
under non-photometric conditions (\ie, the Plaskett; Bond; and non-cmr2, non-wkg
INT images), to the same magnitude system as the rest.  From the results of this
analysis of our own CCD images, we extracted 1,041 well-observed stars with
(a)~at least five measurements {\it and\/} an estimated standard error of the
mean magnitude not worse than 0.1$\,$mag in all of $B$, $V$, $R$, and $I$, and
(b)~a modified Welch-Stetson variability index (Welch \& Stetson 1993) not
larger than 2.5 times the median value.  

Next, the calibrated magnitudes for these 1,041 stars were averaged with the
five previous independent photometric samples ({\it viz.}, Sandage
photoelectric, Eggen ``E'', Eggen ``S'',  Sarajedini, and Platais) after
application of the previously determined offsets to the previous data, so they
should now all be on the same photometric system.  Once known variable stars and
the blended star I-18 were excluded, from this stage there emerged 1,722 stars
which had satisfactory ($\sigma < 0.10\,$mag) measurements in each of $B$ and
$V$ from at least two different studies.  These stars should now represent just
about the best reference list that we can achieve---a gold standard---for
determining and removing systematic photometric offsets from the remaining six
datasets, which did not previously have fully independent calibrations:  the
Sandage (1962) photographic data, and the Sharov (1965), Cannon (1968), McClure
\& Twarog (1977), Caputo \etal\ (1990), and Dinescu \etal\ (1996) results. 
There is no need to impose a selection on the $U$, $R$, and $I$ filters at this
point, since none of these studies presented $U$, $R$, or $I$ data.  The offsets
we derive for these samples are listed in Table~4, but several of the datasets
need a bit more discussion.  

{\bf Sandage photographic:}  Photographic magitudes are listed for six stars
which we could not find on the published charts:  I-123, I-125, I-126, I-127,
I-159, and II-33.  None of the other studies that employed Sandage designations
(Sharov, Eggen \& Sandage, McClure \& Twarog, Caputo \etal, and Platais \etal)
includes any entries for these stars, implying that they couldn't find them
either.  Three other stars, I-158, II-178, and II-205, were manually removed
from the comparison because their magnitudes differed from our adopted reference
magnitudes by more than 1$\,$mag in one or both filters.

{\bf Sharov:} Sharov star 93 had a magnitude residual a bit larger
than 1$\,$mag in $V$.  This star was manually removed from the comparison.

{\bf McClure:}  Photometry is provided for stars identified as 125, 136, and
137, but we were unable to locate these stars on the published finding chart.
The tabulated $V$ magnitudes for stars 173 and 174 are 11.94 and 12.55, 
respectively, and for both the \bmv\ color is given as --0.28.  This is
clearly a mistake.  The two stars are blended together, but it is evident from
McClure \& Twarog's finding chart and other available images that each
component of the blend is fainter than the nearby star 140, for which the
authors list $V = 14.81$.  Our own $V$ magnitudes for stars 173 and 174 are
15.48 and 16.69, while for 140 we get $V = 14.80$.  We exclude the McClure
values for 173 and 174 from further consideration.

{\bf Caputo:}  Three distinct but partially overlapping CCD fields were studied
in \ngc{188}.  As a result, 19 stars in areas of overlap were measured twice
each.  We employed the mean of the two magnitudes in each filter for each of
these stars, but weighted them the same as stars that were measured once since
our purpose here is to compare photometric systems, and the number of
observations for a given star does not affect the system that it is on.  Also,
the Caputo \etal\ study lists the visual magnitude of star II-212 as 15.15.  We
suspect this is a typographical error:  changing the value to 17.15 brings this
observation into excellent agreement with all the other studies that measured
this star.  Finally, the Caputo $B$-band magnitudes for stars I-E3 and I-E18 are
both more than a magnitude fainter than found by Sarajedini, Platais, and our
own study, while the $V$-band magnitudes for both stars are in reasonable
agreement.  These differences are less easy to explain as typographical errors,
so we omit Caputo's measurements of these stars from the rest of our analysis.

{\bf Dinescu:}  Five stars, Dinescu 644, 645, 680, 720, and 734, were
manually removed from the comparison because their magnitudes differed from our
adopted reference magnitudes by more than 1$\,$mag in one or both filters. 

Now, with photometric results from all twelve studies presumably on a common
system, it is our goal to compile a master list with all available results
combined for stars in common.  At this point it becomes useful to consider the
photometric precision of each individual observation.  Our own reductions and
the results of Sarajedini and Platais associate a standard error with each
photometric measurement.  The Sandage (1962) and Eggen \& Sandage (1969)
photoelectric lists, and the Sharov and the Dinescu photographic data do not
include standard errors {\it per se}, but they do specify the number of times
each star was observed.  Caputo \etal\ list only one magnitude in each filter
for most stars, but 19 stars were measured twice because they fell in overlap
regions of different subfields.  Finally, McClure \& Twarog list only
photometric indices on a star-by-star basis, with no indication of precision or
the number of independent measurements for any given star.

For those nine studies which do not list quantitative standard errors, we will
estimate the precision from a comparison with our gold standard.  These
results are summarised in Table~5.  For each study (column 1), columns 2 and 3
give the average root-mean-square difference between the gold-standard values
and the offset-corrected measurements of that study in the $B$ and $V$ filters;
columns 4 and 5 give 1.2533 $\times$ the mean absolute difference; and column 6
gives the arithmetic average of these four estimates of the observational
scatter.  Column 7 gives the mean number of observations per star in the study,
and column 8 gives the inferred standard error of one observation;  for the
first three samples this has been multiplied by a further factor of
$\sqrt{{6\over5}}$ to account for the fact that each of these studies was
included in the definition of the gold-standard system.  For the nine studies
in this table, the uncertainty of an individual tabulated magnitude will be
assumed to be the value in the last column divided by the square root of the
number of times that star was observed.  The same formula will apply without
distinction to the $U$, $B$, and $V$ bandpasses.  This approach is not strictly
rigorous in a mathematical sense, but it should be good enough for our purposes
given the limitations of the data.  

The last three studies are those of Sarajedini, Platais, and us.  In the case of
the Platais study, their data table lists a ``standard deviation'' for each
photometric index; we assume that this represents the standard error of one
measurement because it varies with magnitude but is almost completely
independent of the number of times the star was measured.  Accordingly, we will
divide this standard deviation by the square root of the number of measurements.
Furthermore, the Platais dataset lists $\sigma(V)$ and $\sigma(\hbox{\bmv})$;
for our purposes we take $\sigma^2(B) = {1\over2}\bigl[\sigma^2(V) +
\sigma^2(\hbox{\bmv})\bigr]$.  The Sarajedini data and our own analysis give
uncertainties for the individual magnitudes rather than for derived colors, and
these are what we will use.  However, for all three studies---Sarajedini,
Platais, and ours---we add a further uncertainty of 0.015$\,$mag in quadrature
to lessen the impact of individual measured magnitudes that, due to random
statistical fluctuations and roundoff effects, have tabulated standard errors
$\sim\,$0.000\mag.  (If the listed uncertainties were taken literally, a star
with a standard error of 0.001, for instance, should have four times the weight
of one with a standard error of 0.002.  We believe this would be unrealistic.)
This additional error component can be taken to represent the irreducible
star-to-star photometric differences due to filter mismatch.  In the case of our
own photometric study, this is likely to be unduly pessimistic since we
combined data from six observing runs on three telescopes employing six
different detectors, which should beat down the net effects of bandpass
mismatch.  

Given these precepts, the data from the twelve different sets were merged
into a single star list with weighted mean magnitude values 
in each filter for which observations were available.  The weights were 
the inverse square of the observational uncertainties as laid out in Table~5
and the previous paragraph, and the uncertainty of a final averaged magnitude
was taken to be the inverse square root of the sum of the weights.  The actual
study-to-study repeatability of the results for any individual star was not
considered, except insofar as observations discrepant by more than one magnitude
were discarded.  The information posted at our web site will allow interested
readers to recover the various studies' separate results for any given star.

We would like to mention a couple of small items that came to our attention as
we were working with these datasets.  We present them here in case they
could help future researchers.  Kaluzny \& Shara (1987) identify variable star
V8 as Sandage star III-9; we believe that it is actually III-89.  Zhang \etal\
(2002) have east and west reversed in their Fig.~1.  They also did not note that
their V12 = Sandage III-51.  

\section{PROPER MOTIONS}

Four previous studies have used measurements of stellar proper motions to define
membership indicators for stars in the \ngc{188} field:  Cannon (1968); Upgren,
Mesrobian \& Kerridge (1972); Dinescu \etal\ (1996); and Platais \etal\ (2003). 
The last three studies, in particular, published quantitative membership
probabilities based on each star's proper motion relative to the average cluster
and field proper-motion distributions.  The Dinescu paper also presented a
second probability estimate which includes consideration of the star's position
on the sky as well as its proper motion, but we will not employ this here in
order to simplify the comparison with the other studies.  The different sets of
membership probabilities are generally in good, although not perfect agreement. 
Fig.~1, for example, shows the membership probabilities from Dinescu
plotted against those of Platais for 345 stars in common.  The Dinescu paper
lists photometric and astrometric results only for ``Probable members and stars
of special interest,'' so this plot contains very few stars with Dinescu
membership probabilities $<\,$60\%.  Among these, however, it is noteworthy that
there are many stars with Dinescu probabilities close to 100\% and Platais
probabilities close to 0\%.  Between these extremes there is no perceptible
correlation between the membership determinations apart from the general
statement that there are a few stars for which both studies find the membership
evidence inconclusive.  Finally, there are a very few stars which the Dinescu
study considers to be non-members but nevertheless ``of interest,'' which the
Platais study finds to be members.  We have no way of guessing which other stars
Dinescu \etal\ may have measured and determined to be highly probable
non-members, so if a star is not included in their data table we must assume
that its membership is undetermined by Dinescu \etal, rather than assuming that
they have established non-membership.

The Cannon study of proper motions does not present membership probabilities
{\it per se\/} but rather gives a ratio of the amount by which the motion
of a given star differs from the presumed cluster mean motion, divided by
the uncertainty of the measurement.  In Fig.~2 we plot Cannon's ratio
$\chi$ against the arithmetic mean of the membership probabilities for
stars that appear in one or more of the Upgren, Dinescu, and Platais studies.
The results are more or less as expected in the sense that stars for which
Cannon finds a proper motion close to the cluster mean tend to have
membership probabilities which are high in the other studies.  Conversely,
when Cannon has measured a large motion for a star, the other studies are
emphatic that it is a non-member.  It is curious that, although there are
stars for which Upgren, Dinescu, and Platais alone or in combination find
membership probabilities of 90\%--100\%, none of these stars was included
in Cannon's sample.  Perhaps Cannon's stars were simply too bright for
reliable astrometry in the more recent investigations.  The correlation
between Cannon's $\chi$ and the other studies' membership probabilities
is not perfect: even for stars with Cannon's lowest $\chi$ values there
is a considerable range in the membership probabilities found by others.
Specifically, we find for stars with $0 \leq \chi < 5$, the arithmetic mean
membership probabilities from the other studies is 53\% (based on 84 stars); for
$5 \leq \chi < 15$ the mean membership probability is 22\% (52 stars); for $15
\leq \chi < 30$ it is 6\% (43 stars); and for $\chi \geq 30$ it is 0\%
(112 stars).  Therefore, in order to convert Cannon's $\chi$ to quantities
which can be averaged with the membership probabilities from the other
studies, we assign these numerical probabilities within the stated bins, except
that we convert $\chi < 5$ to a membership probability of precisely 50\%.
That way, if we divide a sample of ``members'' from ``non-members'' by
making a cut at 50\%, Cannon's measurements will not change the decision
one way or the other if $\chi < 5$ and the star has been included in 
{\it any\/} of the other studies.  If Cannon's $\chi \geq 5$, then it
will have the potential to represent a tie-breaking vote against membership if
the other studies are inconclusive among themselves.

\section{DEFINING THE FIDUCIAL SEQUENCES}

Fig.~3 presents a (\bmi,$V$) color-magnitude diagram for probable \ngc{188}
members based upon all the photometric and astrometric data discussed in the
previous sections.  Here we have plotted only stars for which the photometric
uncertainty $\sigma(\hbox{\bmi}) < 0.10\,$mag; large symbols have been used
for stars with measured membership probabilities $\geq 50$\%, and small symbols
have been used for stars without measured proper motions; stars with membership
probabilities determined to be $<\,$50\%, on average among the astrometric
studies, have not been plotted.  (We note that nobody ever assigns a membership
probability of 100\%:  the highest membership probability listed in any of the
studies is 99\%.  Therefore, if a star appears in two proper-motion studies
with a membership probability of 99\% in one and 0\% in the other---the most
extreme case of disagreement possible---such a star will have a mean membership
probability of 49.5\% and will be excluded from the sample of probable members.)
The cluster turnoff and subgiant branch are quite well defined, and a population
of blue stragglers and a binary-star main sequence are also clear.  The
color-magnitude diagram shows the appearance of a ``subdwarf sequence''
consisting of roughly a dozen stars either a magnitude fainter or 0.4$\,$mag
bluer than the principal main sequence.  However, if we apply the more stringent
selection criteria $\sigma(\hbox{\bmi}) < 0.05\,$mag and membership probability
$>\,$90\%, none of these stars survives the cut.  Therefore they do not likely
represent a challenge to our understanding of \ngc{188}'s stellar population.  

Eggen \& Sandage (1969) found a gap roughly a tenth of a magnitude wide
containing no cluster stars just below the main-sequence turnoff of \ngc{188},
centered at magnitude $V=15.55$.  McClure \& Twarog failed to reproduce the gap
with their new photographic data, but nevertheless presented arguments
supporting its reality.  In particular, a gap near this magnitude was found when
only the data from the innermost zone of the cluster were considered.  However,
such a gap is not predicted by modern stellar evolution models if we have
correctly judged the chemical abundances and age of the cluster, even when
diffusive effects are taken into account (\eg, Michaud \etal\ 2004).  Even in
1977, McClure \& Twarog considered the gap to be ``only marginally expected'' on
the basis of then-current theory.  The inset in Fig.~3 shows an enlargement of
the main-sequence turnoff region of \ngc{188}.  The lower of the two arrows
marks $V=15.55$, which is the center of the Eggen \& Sandage gap.  The present
data clearly do not support the presence of a gap at this apparent magnitude. 
The upper arrow, marking a gap which does appear in these data, lies at
$V=15.14$.  McClure \& Twarog also found a gap near this magnitude in their
photometry for stars in Sandage's ring II but not for ring I.  We are not
prepared at present to argue that this gap is real and requires modifying
current stellar evolution models.  However, the anonymous referee of the first
submitted draft of this paper has raised an interesting question: is the
apparent color offset between the brightest stars below the gap and the faintest
stars above it real?  Morphologically, this sort of jog would be expected
if the theoretical models did predict a gap for \ngc{188}.  However, the
referee's question can probably only be answered with better photometry
than is currently available.  The referee has also made the interesting point
that the Eggen-Sandage gap does appear to coincide with a lack of stars on the
binary main sequence in our data.  If so, it is difficult to see how it could
be a consequence of stellar evolution.  Rather, it is almost certainly a
fluke of small-number statistics.

To define the cluster fiducial sequences, we began by printing five diagrams for
probable cluster members and stars without membership determinations: $V$ versus
\bmv, $V$ versus \vmi, $V$ versus \bmi, \bmv\ versus \vmi\ for stars fainter
than $V=15.5$ (mostly member dwarfs), and \bmv\ versus \vmi\ for stars brighter
than $V=15.5$ (mostly member subgiants and giants).  Into each of these five
diagrams a fiducial sequence was sketched by hand, and the locations of
representative points along that sequence were measured with a ruler.  Each of
the five diagrams was measured twice, in orthogonal directions.  That is to say,
consider the $V$ versus \bmi\ plot:  when the color position of the
hand-sketched locus is measured at equally spaced intervals of $V$, we can say
that we have measured the dependent variable \bmi($V$) as a function of the
independent variable $V$; conversely, when the magnitude position of the locus
is measured at fixed intervals of color, we can say that we have measured the
dependent variable $V$(\bmi) as a function of the independent variable \bmi. 
Measurements were made at constant intervals of 0.5$\,$mag in brightness when
$V$ was the independent variable, 0.10$\,$mag when \vmi\ or \bmi\ was the
independent variable, and 0.05$\,$mag when \bmv\ was independent.

When all five plots have been measured in the two orthogonal directions,
we now have five different ways to define a $V$ versus \bmi\ locus:
we can plot $V$ versus \bmi($V$); we can plot $V$(\bmi) versus
\bmi; we can plot $V$ versus \bmv($V$) + \vmi($V$); we can plot $V$(\bmv)
versus \bmv + \vmi(\bmv); and we can plot $V$(\vmi) versus \bmv(\vmi) + \vmi.
We hope that by combining these different approaches to defining the
mean cluster locus, we can beat down the random error in our placement
of the pencil line through the swarm of data points, any systematic bias in our
placement of the millimeter ruler with respect to the plot tickmarks, or any
random or systematic error in the way that we judge the intersection of the
pencil line with the edge of the ruler.  Fig.~4 shows these five different
versions of the $V$ versus \bmi\ color-magnitude diagram, where different
symbols are used to designate the different definitions of the fiducial locus. 
It is apparent that the five different curves are in excellent agreement
except for some slight ambiguity at the top of the giant branch.  

We then printed a new copy of Fig.~4 at twice the scale used for the previous
plots, sketched a smooth curve through the normal points, and read out
compromise representative positions along the curve, again in both directions,
this time at 0.25$\,$mag intervals in $V$ and 0.05$\,$mag in \bmi.  This was
subsequently checked by overplotting the resulting curve on a color-magnitude
diagram for the individual stars, and modest adjustments were made.  We regard
this fiducial sequence as the fundamental result of the present paper.  The
\bmi\ color is attractive for this purpose because it provides the greatest
possible ratio of range to measuring error; it is therefore optimally sensitive
to temperature among the photometric bandpasses available to us at this time. 
This sensitivity also has the effect of providing the greatest possible
discrimination between single main-sequence stars and the unresolved binaries
that lie above and on the red side of the main sequence.  The \bmi\ color has
the additional desirable quality that it is very nearly statistically
independent of $V$, given the way we that have calibrated the raw measurements. 
Therefore there is little or no correlation between random temperature errors
and random luminosity errors (see, for instance, McClure \& Tinsley 1976 to see
why this can be relevant).  

Having produced a smoothed $V$ versus \bmi\ fiducial curve, we then plotted {\it
color\/} versus \bmi\ diagrams for probable member stars and stars of
undetermined membership having photometric color uncertainties less than
0.10$\,$mag, where ``{\it color\/}'' in this case stands for \umb, \bmv, \vmr,
and \vmi, each in turn.  Separate \bmi-{\it color\/} plots were produced for
stars respectively brighter and fainter than $V=15.5$ so as not to confuse
member dwarfs with member giants.  The distinction turns out to be important for
colors involving the $V$ filter, in which dwarfs and giants show perceptibly
different behavior (as Stetson \etal\ 2003 also found in
\ngc{6791}, in the case of the \vmi\ color).  We also produced  $V$ versus {\it
color\/} plots for each of the four colors and measured them in both directions.
The fiducial points listed in Table~6 represent a compromise between values
visually measured in the {\it color-V\/} color-magnitude and \bmi-{\it color\/}
color-color diagrams.  At the end of this process, there remains a certain
amount of jitter in the tabulated curves---due mostly to differences of
perception in the two orthogonal directions---which is probably a fair
representation of the random uncertainties in the adopted fiducial points.

\section{COLOR-COLOR DIAGRAMS}

The top panel of Fig.~5 illustrates the \vmr\ versus \bmv\ color-color diagram
for \ngc{188}.  Here we have plotted only the 796 stars with at least a 50\%
membership probability and estimated color uncertainties $< 0.10\,$mag; stars
with membership probabilities measured to be less than 50\% and stars without
membership determinations have been omitted.  Stars with $V$-band magnitudes
$\geq 15.5$ have been plotted as crosses and those brighter than 15.5 have been
plotted as open circles.  The middle panel of the figure is the analogous plot
for \vmi\ versus \bmv.  As Stetson \etal\ (2003) found in the case of the \vmi\
versus \bmv\ color-color diagram for \ngc{6791}, there is a clear separation of
cluster giants and dwarfs for the reddest colors, in the sense that at long
wavelengths giants are bluer than dwarfs for fixed \bmv\ colors.  There is no
clear separation with luminosity in an \rmi\ versus \bmi\ plot, which
establishes an extra absorption component in the $V$ bandpass at high surface
gravity as the most likely cause of the difference.  The bottom panel of Fig.~5
plots the sum of the two long-wavelength colors versus \bmv; combining the
long-wavelength colors produces a slight improvement in the ratio of separation
to scatter.  The solid line in this plot represents the equation (\vmr)+(\vmi) =
1.70(\bmv)--0.04, which we will use later as a discriminator between giants and
dwarfs.  This criterion is probably useful only for \bmv\ colors larger than
0.90--1.00 (or \bmi$\,>\,$1.9--2.1), but if a bluer star is found to be lying
significantly above this line, it probably has large photometric errors or an
unusual spectral-energy distribution.  

This giant/dwarf discriminator can be used as a (weak) membership indicator
for those stars where the proper-motion evidence is marginal or lacking.  The
upper panel of Fig.~6 is the (\vmr)+(\vmi) versus \bmv\ diagram for the
163 stars in the \ngc{188}\ field with measured proper motions that imply
membership probabilities $10\% \leq p \leq 49\%$, and photometric uncertainties
in (\vmr)+(\vmi) $< 0.10\,$mag.  Here we have plotted stars lying above the
giant/dwarf dividing line (probable giants if the stars are red, possible
photometric mistakes or peculiar stars if they are blue) as open circles, and
stars below the line (probable normal dwarfs) as crosses.  
The lower panel of this figure shows the $V$ versus \bmi\ color-magnitude
diagram for the same stars, where the symbols have the same meaning as in the
upper panel and the solid curve is our adopted cluster fiducial sequence.  Here
we see that many stars with ambiguous proper-motion membership determinations
fall very near \ngc{188}'s main sequence and have the photometric properties of
dwarfs.  It is likely that the membership probabilities of these stars have been
estimated low because their proper motions fall in the tail of the random
distribution of measuring errors.  This is a consequence of utilizing a fixed
cluster/field star ratio in the probability calculations.  If, for instance, the
astrometrists were to consider the cluster/field star ratio only for those stars
which fell within a few standard deviations of the mean cluster photometric
locus, that ratio would be much higher than the ratio for all stars considered
together.  With such a calculation the membership probabilities of these stars
would be significantly increased.  (Of course, using one cluster/field ratio
for stars near the principal locus and another for stars off of it would
greatly reduce the likelihood of finding peculiar cluster members.  Our point
here is not to criticize the standard method of membership determination; it
is merely to point out that adopting any arbitrary membership cutoff larger than
0\% involves the near certainty of rejecting some genuine cluster members. 
Stars with intermediate membership probabilities---especially those with
photometric properties consistent with cluster membership---should be considered
possible members at least until other evidence, such as radial velocity or
chemical abundances, can be included in the analysis.)

There is one star with the photometric properties of a giant that lies very near
the cluster giant branch.  This is Sharov~61 (=~Cannon~661 = Sarajedini~594 =
Platais~4856 = our~8648) with an average proper-motion membership probability
$p\,=\,11$\%.  There are also two stars with giant-like colors that lie well to
the red of the cluster giant branch:  the bluer of the two is Sandage~III-94 (=
Sharov~78 = Cannon~623 = Upgren~200 = Sarajedini~254 = Platais~5607 = our 10120)
with $p\,=\,45$\% \footnote{Table~3 of Dinescu \etal\ also identifies this star
as 1282 in their numbering scheme, but no data are provided for the star in
their Table~2.  This probably means that they, too, considered it to have a
membership probability $<60$\%.}, and the other is Sandage~C (= Sharov~26 =
Cannon~672 = Platais~4460 = Sarajedini~1086 = our~6522) with $p = 25$\%.  In the
direction of \ngc{188} red field stars with giant-like photometric properties
are quite rare, as can be seen in Fig.~7, where we have produced the identical
diagrams for 1,971 stars with measured proper-motion membership probabilities
$\leq\,9$\%.  Although this diagram contains twelve times as many stars as
Fig.~6, there are {\it no\/} field stars redder than \bmi$\,=\,$2.2 with the
photometric properties of a giant down at least to $V\,=\,20$.  Therefore, in
adopting a fixed (cluster/field) ratio for their probability calculations
regardless of color, magnitude, or other independent information, the
proper-motion studies may well have underestimated the membership probabilities
of these two stars.  A quick search of the SIMBAD database produces no
spectroscopic information for either star, but they probably do deserve
spectroscopic study.  If they are cluster members, they lie in the general
region where carbon-, barium-, and related stars can be found.

Figs.~6 and 7 both show stars in the blue-straggler zone of the color-magnitude
diagram in roughly similar numbers, although the latter diagram contains twelve
times as many stars overall as the former.  Again, this may suggest that the
membership probabilities of at least some of these stars have been
underestimated because the number of true field stars in this box of magnitude
and color is rather low compared to the rest of the color-magnitude diagram.

Fig.~8 shows the same two diagrams for 140 stars with adequate photometric
data but whose proper motions have not been measured.  Since to have been
omitted from the astrometric studies a star is likely to be either faint or
crowded, the precision of these photometric indices tends to be poor.  As a
result, the giant/dwarf discriminator is noisy and not very effective. 
Nevertheless, there is no evidence that any possible giant members of the
cluster have been missed by the astrometric studies, although there are a few
additional faint stars that might belong to the cluster main sequence.

Fig.~9 compares our fiducial sequence to Landolt's photometric standards---which
may be taken at least as a representative, if not random, sample of Solar
Neighborhood stars dominated by the local Population~I---in the \umb, \bmv,
\vmr, and \vmi\ colors, each plotted against \bmi.  In each of the four panels,
stars identified by Smith \etal\ (2002) as belonging to luminosity class III are
shown as large filled circles, stars belonging to class V are large crosses
and stars of other or undetermined luminosity classes are represented by small
crosses.  Here we see that giants and dwarfs do indeed tend to be separated
in diagrams involving the $V$ magnitude.  We also see that the agreement between
the cluster sequence and the field stars is quite good in every panel except the
one showing \vmr, where our cluster-star \vmr\ colors are about 0.015\mag\ to
0.025\mag\ redder than the field-star colors for fixed \bmi.  Given that the
differences in the reddening and metallicity between \ngc{188} and the average
field star are not likely to be substantial, and especially in view of the fact
that such an offset appears in only the one panel (of all these colors, \vmr\ is
the one {\it least\/} affected by reddening and line blanketing), our adopted
$R$ magnitude scale is probably $\sim\,$0.02\mag\ too bright, \ie, the
$R$ magnitudes are quantitatively too small so the \vmr\ colors are too large. 
As the $R$ magnitude zero-point was determined from only two independent
studies, each of which is probably uncertain at a level of $\sim\,$0.03\mag, and
which differ between themselves by 0.09\mag\ (Table~3), such a level of
systematic error is entirely plausible.  Notice that if this inference is 
correct, then Sarajedini \etal\ measured their $R$ magnitudes only 0.02$\,$mag
too faint, rather than the 0.04$\,$mag given in Table~3; this is similar to
their offsets in the $V$ and $I$ filters.

Fig.~10 presents a comparison of the fiducial sequences of three of the
most famous old open clusters, \ngc{188} (crosses; present study), M$\,$67
(closed circles), and \ngc{6791} (open circles); these latter two sequences were
adapted by Sandage \etal\ (2003) from previous photometric
investigations by Montgomery, Marschall \& Janes (1993; M67) and Stetson \etal\
(2003; \ngc{6791}).  Comparing the latter two clusters to an \ngc{188}\ fiducial
derived from unpublished photometry by VandenBerg \& McClure (a
preliminary analysis of the DAO images forming a part of the present study),
Sandage \etal\ found that the lower main sequence of \ngc{188}\ did not
appear to lie parallel to those of the other two clusters.  Conventional
stellar-structure theory would have a very difficult time explaining this
circumstance if the metal abundances and foreground reddening values adopted
for these systems were even remotely correct.  The mean main sequence obtained
in this paper from an appreciably expanded body of observational data does
not confirm the disturbing trend found by Sandage \etal\ \  In fact, not only
does the \ngc{188} sequence now parallel the other two, it also closely overlies
the unevolved main sequence of M$\,$67, implying that either we have correctly
evaluated the relative metal abundances, reddenings, {\it and\/} distance 
moduli of these two clusters, or there is a pernicious conspiracy of systematic
errors in two or more of these quantities.  The unevolved main sequence of
\ngc{6791}\ stands well to the red of the sequences for the other two clusters
by an amount which is consistent---as shown by Sandage \etal---with an
overall metal abundance which is greater by about +0.4$\,$dex than the
common abundance of M$\,$67 and \ngc{188}.\footnote{As discussed by Sandage
\etal, there is considerable evidence that the reddening of \ngc{6791}\ may
be closer to $E(\bmv) = 0.15\,$mag than to the value $E(\bmv) \approx
0.10\,$mag found by Stetson \etal\ (2003).  If the lower reddening were adopted
it would be necessary to make a corresponding adjustment to the apparent
distance modulus of the cluster---to something like 13.25---in order to bring
its lower main sequence into coincidence with the red edge of the Hipparcos
color-magnitude diagram for nearby stars.  The resulting comparison with M$\,$67
and \ngc{188}\ would still be very similar to that shown in Fig.~10.}

Finally, Fig.~11 shows a comparison between our fiducial main sequences in
\ngc{188}\ and those of the local field population as defined by
the stars in the Gliese (1969) {\it Catalogue of Nearby Stars\/} having
photometry by Bessell (1990) and Hipparcos parallaxes (Perryman \etal\ 1998)
that are at least ten times their standard errors.  In the $(\bmv)_0$--$M_V$
(top panel) and $(\vmi)_0$--$M_V$ (bottom panel) color-magnitude diagrams the
agreement is excellent, as would be expected if (a)~the local field population
and \ngc{188} have similar chemical abundances (both are estimated to have close
to Solar abundance patterns), (b)~we have correctly judged the reddening and
apparent distance modulus of the cluster, {\it and\/} (c)~the systematic errors
in our photometry are small.  Of course, as before, a conspiracy of errors in
two or more of these assertions could produce spurious agreement.  It is evident
that the agreement in the $(\vmr)_0$--$M_V$ diagram (middle panel) is
perceptibly worse.  Since, as we've stated before, \vmr\ is the one color out of
all these that is {\it least\/} sensitive to both metallicity and reddening
effects, it would be hard to remove this discrepancy by altering one or both of
these quantities, especially given the good agreement in $(\bmv)_0$ and
$(\vmi)_0$.  We believe that this diagram is additional evidence for a
systematic error in the zero-point of our adopted $R$ magnitude scale, in the
sense that we have measured \ngc{188}\ stars too bright in $R$, and therefore
too red in \vmr, by about 0.02$\,$mag.  We have {\it not\/} removed this offset
from the cluster data which we have posted on the Internet; users of these data
for future research should be aware of this fact.

\section{Summary}

We have obtained new CCD photometry for the old open cluster \ngc{188}\ with
the Dominion Astrophysical Observatory 1.8m telescope.  Because these data
are difficult to calibrate we have found it useful to consider additional
data donated by Howard Bond, and data obtained from the Isaac Newton Group
archive, as well as all the previous cluster photometry that we were able
to locate in the literature.  Direct star-by-star comparisons have been
undertaken to determine and remove the systematic differences in photometric
zero-point that are found among the various studies.  At the same time
we have merged the independent membership indices that have been produced
by four different astrometric studies of the cluster field and matched
those membership probabilities with the photometry for the same stars.  

It has been our aim to combine all the available datasets for \ngc{188}, not to
critique them.  It must be remembered that we have removed only differences of
photometric zero-point among the various published studies of the cluster.  In
actuality, systematic calibration errors can be more complicated than this:
magnitude scales may vary with the color of the star if bandpass mismatch is not
adequately modeled, or with the brightness of the star if the detector is not
accurately linear, or with the position of the star on the sky when imaging
detectors have significant distortions, aberrations, or scattered-light
problems.  The systematic errors between one dataset and another can in
principle be modeled as a Taylor expansion in these independent variables.  We
have contented ourselves with estimating and removing only the zeroth-order
terms in these Taylor expansions.  To have attempted more than this would have
required appreciably more work, and also it would likely have involved choosing
one of the data sets to be the ``correct'' one and adjusting all the others to
match it.  By considering only mean values, it has been straightforward to
define an ``average'' photometric system without having to choose any one study
as being more authoritative than the others.

As a consequence, any given study has been accurately referenced to the average
of all studies only for stars near the average color, magnitude, and position;
stars at extreme values of these variables may retain systematic differences
from one study to another.  We contend that these introduce mostly noise and not
systematic error to the final data products, because---once the zero-point
offsets have been removed---any study which has positive systematic errors in
some regions of color-magnitude-position space will have comparable negative
systematic errors in other parts of that space.  Furthermore, there is little
reason to expect that such systematic errors will be identical in the various
studies (with the exception of the photographic studies that were calibrated on
the basis of the photoelectric studies: in this situation any color- or
magnitude-dependent errors in the photoelectric work may well have been mapped
into the photographic results).  Average photometric indices determined from
multiple investigations should therefore be more reliable than results for a
star studied only once.  

By adopting a single value for the photometric uncertainty inherent to each
study, regardless of the brightness of the star or the filter used, we have
again accounted for the different precisions of the various datasets only to
zeroth order.  Once again, we think this is a reasonable compromise, and that a
more complex approach to the weighting of the data would add extra work out of
proportion to the improvement achieved.  We believe that we have ``done no
harm'' to the data in the Hippocratic sense of the phrase: our homogenized results
for any given star are probably not worse than the results from the best of the
studies in which that star was included.  Nevertheless, the reader is encouraged
to remember the compromises that we have made if using our results for future
research.  It would be unwise to base critical scientific conclusions on the
photometric indices or membership status of any given star, especially if that
star was observed only once or a few times.  Conclusions based upon the bulk
properties of the stars in \ngc{188} as they have been presented here are
probably somewhat safer.

At our web site\footnote{http://cadcwww.hia.nrc.ca/stetson/} we have posted
ASCII data files containing (a)~our final merged photometry, membership
information, and J2000.0 equatorial coordinates for 9,228 stars in the field of
NGC188, based upon all available studies; (b)~our derived photometry for 4,863
stars based upon our own analysis of 299 CCD images; (c)~a transit table
relating our sequential identification numbers to the star identifications
employed in all the previous studies that we have considered; and (d)~those data
tables from previous published studies which we have entered into the computer
by hand, because we were unable to locate electronic copies.  These files
should make it easy for interested readers to reanalyze these data if our
approach seems inadequate for their purposes.

\bigskip

The authors would like to thank Ata~Sarajedini and Imants~Platais for providing
machine-readable copies of their data tables.  PBS continues to be very
grateful to Howard E. Bond, the Canadian Astronomy Data Centre, and the Isaac
Newton Group Archive for providing data for this and other ongoing studies. 
This work was supported, in part, by a Discovery Grant to DAV from the Natural
Sciences and Engineering Research Council of Canada.

\clearpage

\figcaption[stetsonfig1.eps]{A comparison of quantitative proper-motion
membership probabilities measured by Dinescu \etal\ (vertical axis) to those
measured by Platais \etal\ (horizontal axis).  This plot illustrates the
imperfect agreement between different membership studies.  Dinescu \etal\
actually published final tabulated values for stars which they considered to
have membership probabilities $>\,$60\%, plus a few other ``interesting'' stars. 
This explains the sparseness of points in the lower part of the figure.  The
actual densities of stars at the extreme left and right of the diagram may not
be as they appear because of the custom of publishing membership probabilities
as integer percentiles.  So, for instance, the point at (99,99) could actually
contain any number of stars overlying one another.  Nevertheless, even among the
stars that Dinescu \etal\ considered probable members, it is evident that
Platais \etal\ find a number of stars they consider to be certain nonmembers. 
There are also at least two stars which Dinescu \etal\ consider to be
interesting nonmembers, that Platais \etal\ find to be almost certain members.}

\figcaption[stetsonfig2.eps]{A comparison of Cannon's proper-motion statistic,
$\chi$, to the average of the numerical membership probabilities determined from
three other astrometric studies for stars in common.  $\chi$ basically
represents the ratio of a star's measured proper motion relative to the adopted
mean cluster motion, to the estimated standard error of the proper-motion
measurement.  It is clear that while the one-to-one correspendence between
$\chi$ and the quantitative membership probabilities is not good, nevertheless
there is a general trend in the sense that the {\it average\/} probability of
membership tends downward as $\chi$ increases.  For our purposes, we have simply
divided Cannon's $\chi$ into four ranges, as indicated by dashed lines, and have
assigned numerical membership probabilities of 50\% (the actual arithmetic mean
probability in this zone is 53\%), 22\%, 6\%, and 0\% for stars lying in these
four zones, from left to right respectively.  The tendency for the numerical
membership probabilities to be quantized is probably related to the fact,
illustrated in Fig.~1, that the methodology tends to favor membership indices
strongly concentrated toward 99\% and 0\%.  When two or more studies differ, the
mean membership probability tends toward averages of these two
numbers.} 

\figcaption[stetsonfig3.eps]{A $V$ versus \bmi\ color-magnitude diagram for
stars in the field of the open cluster \ngc{188}.  Only stars with photometric
uncertainty $\sigma(\bmi) < 0.10\,$mag are plotted, and stars with measured
membership probabilities greater than 50\% are represented by large $\times$'s,
while smaller symbols represent stars whose proper motions have not been
measured. (inset) An enlargement of the main-sequence turnoff of \ngc{188}.
Is there a gap present at $V=15.14$ (upper arrow)?  We invite the reader to be
the judge.  The lower arrow marks $V=15.55$, where Eggen \& Sandage claimed
to find a gap in their data.} 
\figcaption[stetsonfig4.eps]{The mean cluster locus in the $V$ versus \bmi\
color-magnitude diagram, as estimated by eye in five different trials. 
Different symbol types represent different trials.  The agreement is generally
good, with some slight ambiguity near the top of the giant branch.} 

\figcaption{A plot showing the gravity sensitivity of {\it BVRI\/} colors for
stars redder than \bmv$\,\sim\,$0.95:  the upper panel shows \vmr\ plotted
against \bmv, while the middle panel shows \vmi\ plotted against \bmv.
In each case, only stars having mean measured proper-motion membership
probabilities $\geq\,$50\% and $\sigma$(color)$\,<\,$0.10 are shown; stars with
undetermined membership have not been plotted.  Open circles represent stars
brighter than $V=15.5$ (roughly the main-sequnce turnoff) which should be
subgiants and giants, while $\times$'s represent fainter, \ie, main-sequence,
stars.  The bottom panel shows the effect of plotting the sum of \vmr\ and \vmi\
against \bmv: the separation between the giant and main-sequence branches,
relative to the photometric scatter in each, is slightly improved.  The sloping
line represents the equation (\vmr)+(\vmi) = 1.70(\bmv)--0.04, which we will use
as a giant/dwarf discriminator for red stars, and an index of possible stellar
peculiarity or photometric mistakes for blue stars.}

\figcaption[stetsonfig6.eps]{(Upper panel) A plot of (\vmr)+(\vmi) against \bmv\
color for stars in the \ngc{188} field with proper-motion membership
probabilities between 10\% and 50\%.  Open circles are used to represent stars
having (\vmr)+(\vmi)$\,<\,$1.70(\bmv)--0.04 and $\times$'s represent stars lying
below this line.  For stars redder than \bmv = 0.95 or so, which corresponds to
\bmi\ $\gtsim$ 2.0, open circles are likely to be giants and $\times$'s are
likely to be dwarfs. Bluer than this limit normal stars are not expected to
have (\vmr)+(\vmi)$<$1.70(\bmv)--0.04; open circles with \bmv $<$ 0.95 probably
represent peculiar stars or unusually large measuring errors.  (Lower panel)  A
$V$ versus \bmi\ color magnitude diagram for the same stars as are shown in the
upper panel.  Symbol types have the same significance, and the solid curve
represents our adopted fiducial cluster sequence.  This figure shows a striking
concentration of these stars toward the mean cluster locus, especially near the
turnoff and the lower main sequence.  One star with the long-wavelength colors
of a giant lies close to the cluster giant branch, and two others lie in a zone
where few giant stars are found in the surrounding field population (cf.\
Fig.~7).  It is likely that many or most of these stars are actually cluster
members.  A number of stars having marginal membership indices lying in the blue
straggler region could also be actual cluster members.}

\figcaption[stetsonfig7.eps]{(Upper panel) A plot of (\vmr)+(\vmi) against \bmv\
color for stars in the \ngc{188} field with proper-motion membership
probabilities less than 10\%.  Open circles are used to represent stars having
(\vmr)+(\vmi)$\,<\,$1.70(\bmv)--0.04 and $\times$'s represent stars lying below
this line.  For stars redder than \bmv = 0.95 or so, which corresponds to \bmi\ 
$\gtsim$ 2.0, open circles are likely to be giants and $\times$'s are likely to
be dwarfs. Bluer than this limit normal stars are not expected to have
(\vmr)+(\vmi)$<$1.70(\bmv)--0.04; open circles with \bmv $<$ 0.95 therefore
probably represent peculiar stars or unusually large measuring errors.  The
solid curve represents our fiducial sequence for \ngc{188}.  (Lower panel)  A
$V$ versus \bmi\ color magnitude diagram for the same stars as are shown in the
upper panel.  Symbol types have the same significance, and the solid curve
represents our adopted mean cluster locus.  While some of the stars with low
membership probabilities {\it could\/} belong to the cluster, there is no
pronounced concentration of stars toward the fiducial cluster sequence.  Rather,
the impression is of a uniform sheet of stars spread across the diagram.  Note
in particular the almost complete absence of stars with the colors of giants
(open circles) at {\it any\/} magnitude with colors redder than
\bmi$\,\sim\,$2.0: in this color range dwarfs appear to provide almost all of
the field-star counts.  This emphasizes the abnormality of the two stars Sandage
III-94 and Sandage~C, seeming giants which lie in an otherwise very sparsely
populated region of the color-magnitude diagram.}

\figcaption[stetsonfig8.eps]{(Upper panel) A plot of (\vmr)+(\vmi) against \bmv\
color for stars in the \ngc{188} field without any proper-motion membership
measurements.  Open circles are used to represent stars having
(\vmr)+(\vmi)$<$1.70(\bmv)--0.04 and $\times$'s represent stars lying below this
line.  For stars redder than \bmv = 0.95 or so, which corresponds to \bmi\
$\gtsim$ 2.0, open circles are likely to be giants and $\times$'s are likely to
be dwarfs. Bluer than this limit normal stars are not expected to have
(\vmr)+(\vmi)$<$1.70(\bmv)--0.04; open circles with \bmv $<$ 0.95 probably
represent peculiar stars or unusually large measuring errors.  The solid curve
represents our fiducial sequence for \ngc{188}.  (Lower panel)  A $V$ versus
\bmi\ color magnitude diagram for the same stars as are shown in the upper
panel.  Symbol types have the same significance, and the solid curve
represents our adopted mean cluster sequence.  Here we see that most of
the stars without proper-motion measurements are faint, but the colors of a
number of them are consistent with lying on the cluster main sequence.  We
note a number of brighter stars also lie near the cluster main sequence, but
about half of them have long-wavelength colors suggesting that they are giants
rather than dwarfs.  At a guess, we would state that probably these stars {\it
are\/} cluster dwarfs but the same crowding that precluded astrometry has
inflated their photometric errors, scattering these stars into the giant zone of
the color-color diagram.}

\figcaption[fig9.eps] {Color-color diagrams relating the conventional colors
\umb\ (top left), \bmv\ (top right), \vmr\ (bottom left), and \vmi\ (bottom
right) to the ultra-sensitive color \bmi, for Landolt's (1992) equatorial
standard stars.  Large filled circles represent stars identified as belonging to
luminosity class III by Smith \etal\ (2002), large $\times$'s represent
luminosity class V stars according to Smith \etal, and small $\times$'s
represent stars that Smith \etal\ assigned to different luminosity classes and
stars not included in their listing.  In each case, the solid curve represents
our fiducial locus for \ngc{188}.  The separation between dwarfs and giants for
$\bmi \gtsim 2$ is evident in colors involving the $V$ filter for both the
cluster and the field.  Our derived colors for \ngc{188} stars agree quite well
with the majority of Landolt's standard stars except in the \vmr\ plot, where
the cluster sequence lies from 0.015$\,$mag to 0.025$\,$mag to the red of the
standard stars.  This is most probably due to an unavoidable systematic error in
the zero-point of our derived $R$-magnitude scale.}

\figcaption[fig10.eps] {A comparison of the present \ngc{188}\ fiducial sequence
(crosses) to previously published fiducial sequences for the old open clusters
M$\,$67 (closed circles) and \ngc{6791} (open circles), after all have been
corrected for reddening and apparent distance modulus by the amounts listed in
the legend.  The \ngc{188}\ sequence published by Sandage \etal\ (2003) was not
parallel to that of M$\,$67 along the lower main sequence, as it should have
been since the two clusters are believed to have the same metal abundance.  The
new fiducial sequence for \ngc{188} is not only parallel to that for M$\,$67, it
also closely overlies it.  This is would be expected if the two clusters have
the same metal abundance {\it and\/} we have correctly estimated the relative
distances and reddenings of the two systems.  The unevolved main sequence of
\ngc{6791} lies appreciably to the red of those of the other two clusters,
consistent with the conclusion that this cluster is significantly more 
metal rich than they are (see, \eg, Sandage \etal\ 2003).}

\figcaption[fig11.eps] {The dereddened color--absolute visual magnitude diagrams
for stars in the Gliese catalog with the smallest relative parallax errors
(crosses) are compared to the fiducial sequences for \ngc{188} (solid curves) in
(top) $(\bmv)_0$, (center) $(\vmr)_0$, and (bottom) $(\vmi)_0$.  Since the
nearby field population and \ngc{188}\ are both expected to have near-Solar
abundance ratios, we would expect the unevolved main sequences in these two
populations to coincide if we have correctly judged the reddening and apparent
distance modulus of the cluster.  We find that this is indeed the case for both
$(\bmv)_0$ and $(\vmi)_0$, while the cluster sequence in $(\vmr)_0$ skims the
upper (red or bright) envelope of the local field main sequence.  We believe that
this is a consequence of a not unreasonable systematic error in the zero-point
of our $R$ magnitude scale, in the sense that the cluster stars have been
measured too bright in $R$ ($\Rightarrow$ too red in \vmr) by about 0.02$\,$mag. 
Shifted bluer (to the left) by 0.02$\,$mag in the center panel (dashed curve,
hardly visible in this plot), the \ngc{188} sequence comes much closer to
threading its way through the middle of the field-star sequence.} 

\baselineskip 0.1 true cm
\footnotesize

\begin{deluxetable}{llrccccl}
\tablecaption{Photometric datasets for \ngc{188}}
\tablecolumns{8}
\tablewidth{0pt}
\tablehead{
\colhead{(1)} & \colhead{(2)} & \colhead{(3)} & \colhead{(4)} & \colhead{(5)} & 
\colhead{(6)} & \colhead{(7)} & \colhead{(8)} \\
\colhead{Source} & \colhead{Type} & \colhead{Stars} &
\colhead{$U_{\hbox{lim}}$} & \colhead{$B_{\hbox{lim}}$} &
\colhead{$V_{\hbox{lim}}$} & \colhead{$I_{\hbox{lim}}$} & \colhead{Calibration
standards} 
}
\startdata
Sandage pe & photoelectric &  123 & 19.1 & 19.8 & 18.7 &  ---  & Johnson \\
Sandage pg & photographic  &  563 &  ---  & 19.0 & 18.0 &  ---  & Sandage pe \\
Sharov     & photographic  &   61 &  ---  & 14.9 & 14.0 &  ---  & Sandage pe \\
Cannon     & photographic  &  291 &  ---  & 15.1 & 14.4 &  ---  & Sandage pe \\
Eggen E    & photoelectric &   75 & 17.5 & 17.4 & 16.7 &  ---  & Johnson \\
Eggen S    & photoelectric &   97 & 19.3 & 20.0 & 18.7 &  ---  & Johnson \\
Upgren     & photographic  &  228 &  ---  &  ---  & 15.2 &  ---  & various \\
McClure    & photographic  &  664 &  ---  & 19.0 & 18.0 &  ---  & Eggen \\
Caputo     & CCD           &  285 &  ---  & 21.5 & 19.9 &  ---  & Eggen \\
Dinescu    & photographic  &  355 &  ---  & 16.6 & 15.9 &  ---  & Sandage pe \\
Sarajedini & CCD           & 1520 & 19.1 & 20.4 & 19.3 & 18.2 & Landolt \\
Platais    & CCD           & 7812 &  ---  & 22.2 & 21.1 &  ---  & Landolt? \\
present    & CCD           & 4863 &  ---  & 23.2 & 22.4 & 21.1 & Stetson (Landolt)\\
\enddata
\end{deluxetable}

\clearpage

\begin{deluxetable}{lrcc}
\footnotesize
\tablecaption{Astrometric comparisons}
\tablecolumns{4}
\tablewidth{0pt}
\tablehead{
\colhead{(1)} & \colhead{(2)} & \colhead{(3)} & \colhead{(4)} \\
\colhead{Sample} & \colhead{Stars} & \colhead{$\sigma(\alpha)$} & \colhead{$\sigma(\delta)$} \\
&& \colhead{$\prime\prime$} & \colhead{$\prime\prime$} 
}
\startdata
USNO GSC   &  6953 & 0.42 & 0.33 \\
DSS1 ``O'' &  7457 & 0.64 & 0.60 \\
DSS2 ``B'' &  8792 & 0.39 & 0.39 \\
DSS2 ``R'' &  9763 & 0.19 & 0.19 \\
DSS2 ``I'' &  7727 & 0.50 & 0.49 \\
Cannon     &   289 & 0.87 & 0.61 \\
Upgren     &   135 & 1.64 & 1.64 \\
Dinescu    &   355 & 0.07 & 0.08 \\
Sarajedini &  1520 & 0.08 & 0.09 \\
Platais    &  7812 & 0.11 & 0.09 \\
present    &  4863 & 0.04 & 0.04 \\
\enddata
\end{deluxetable}

\clearpage

\begin{deluxetable}{lccccc}
\tablecaption{Comparison of photometric studies for \ngc{188}}
\tablecolumns{6}
\tablewidth{0pt}
\tablehead{
\colhead{(1)} & \colhead{(2)} & \colhead{(3)} & \colhead{(4)} & \colhead{(5)} & 
\colhead{(6)} \\
\colhead{Sample} & \colhead{$U$} & \colhead{$B$} & \colhead{$V$} & \colhead{$R$} & \colhead{$I$}
}
\startdata

\multicolumn{6}{c}{1. Mean magnitude offsets (Sarajedini minus other)}\\

Sandage pe           & --0.031  & +0.014 & +0.011 & \nodata & \nodata\\
Eggen E              & +0.064  & +0.066 & +0.080 & \nodata & \nodata\\
Eggen S              & --0.014  & +0.038 & +0.039 & \nodata & \nodata\\
Platais              & \nodata & +0.015 & +0.022 & \nodata & \nodata\\
cmr2                 & \nodata & +0.072 & +0.003 & \nodata & \nodata\\
wkg                  & \nodata & +0.089 & +0.044 & +0.086 & +0.050\\
\\
average              & +0.005  & +0.042 & +0.029 & +0.043 & +0.025\\
standard deviation   &  0.041  &  0.034 &  0.031 &  0.061 &  0.036\\
(incl. Sarajedini)\\
\\
\multicolumn{6}{c}{2. Standard deviations}\\
Sandage pe           &  0.114  &  0.070 &  0.071 & \nodata & \nodata\\
Eggen E              &  0.076  &  0.055 &  0.042 & \nodata & \nodata\\
Eggen S              &  0.072  &  0.066 &  0.056 & \nodata & \nodata\\
Platais              & \nodata &  0.049 &  0.038 & \nodata & \nodata\\
cmr2                 & \nodata &  0.040 &  0.032 & \nodata & \nodata\\
wkg                  & \nodata &  0.078 &  0.099 &  0.083  & 0.080\\
\\
average              &  0.087 &  0.060 &  0.056 &  0.083 &  0.080\\
\\
\\
\\
\\
\multicolumn{6}{c}{3. Median magnitude offsets (Sarajedini minus other)}\\
Sandage pe           & --0.032  & +0.009 & +0.002 & \nodata & \nodata\\
Eggen E              & +0.066  & +0.063 & +0.081 & \nodata & \nodata\\
Eggen S              & --0.018  & +0.032 & +0.035 & \nodata & \nodata\\
Platais              & \nodata & +0.019 & +0.021 & \nodata & \nodata\\
cmr2                 & \nodata & +0.072 & +0.004 & \nodata & \nodata\\
wkg                  & \nodata & +0.082 & +0.036 & +0.087 & +0.060\\
\\
average              & +0.004 & +0.040  & +0.026 & +0.044 & +0.030\\
standard deviation   &  0.043 &  0.033  &  0.029 &  0.062 &  0.042\\
(incl. Sarajedini)\\
\\
\multicolumn{6}{c}{4. $\sqrt{\pi\over2}\,\times\,$mean absolute deviations}\\
Sandage pe           &  0.085  &  0.051 &  0.048 & \nodata & \nodata\\
Eggen E              &  0.077  &  0.052 &  0.041 & \nodata & \nodata\\
Eggen S              &  0.063  &  0.057 &  0.045 & \nodata & \nodata\\
Platais              & \nodata &  0.035 &  0.028 & \nodata & \nodata\\
cmr2                 & \nodata &  0.036 &  0.030 & \nodata & \nodata\\
wkg                  & \nodata &  0.078 &  0.085 &  0.080  & 0.083\\
\\
average              &  0.075  &  0.052 &  0.046 &  0.080  & 0.083\\
\\
\\
\\
\\
\multicolumn{6}{c}{5. Number of stars used in the comparison}\\
Sandage pe           &     80  &    110 &    111 & \nodata & \nodata\\
Eggen E              &     67  &     69 &     69 & \nodata & \nodata\\
Eggen S              &     63  &     89 &     90 & \nodata & \nodata\\
Platais              & \nodata &   1379 &   1506 & \nodata & \nodata\\
cmr2                 & \nodata &    332 &    337 & \nodata & \nodata\\
wkg                  & \nodata &    293 &    303 &    299  &   300\\
\\
\multicolumn{6}{c}{6. Additive correction to be applied}\\
Sandage pe           & --0.036  & --0.029 & --0.022 & \nodata & \nodata\\
Eggen E              & +0.061  & +0.023 & +0.052 & \nodata & \nodata\\
Eggen S              & --0.020  & --0.006 & +0.009 & \nodata & \nodata\\
Sarajedini           & --0.004  & --0.041 & --0.028 & --0.043 & --0.028\\
Platais              & \nodata & --0.024 & --0.006 & \nodata & \nodata\\
cmr2                 & \nodata & +0.031 & --0.024 & \nodata & \nodata\\
wkg                  & \nodata & +0.045 & +0.010 & +0.043 & +0.027\\
\enddata
\end{deluxetable}

\clearpage

\begin{deluxetable}{lcc}
\tablecaption{Comparison of photometric studies for \ngc{188}}
\tablecolumns{3}
\tablewidth{0pt}
\tablehead{
\colhead{(1)} & \colhead{(2)} & \colhead{(3)} \\
\colhead{Sample} & \colhead{$B$} & \colhead{$V$} 
}
\startdata
\multicolumn{3}{c}{Mean offsets (Adopted minus other)}\\

      Sandage pg\ \ \ \ \ \ \ \  & --0.038 & --0.023 \\
      Sharov      & --0.040 & --0.025 \\
      Cannon      & --0.033 & --0.007 \\
      McClure     & +0.003 & +0.034 \\
      Caputo      & --0.024 & +0.033 \\
      Dinescu     & --0.016 & --0.004   \\
      \\
\multicolumn{3}{c}{Standard deviations} \\

      Sandage pg\ \ \ \ \ \ \ \  &  0.109 &  0.078   \\
      Sharov      &  0.047 &  0.125   \\
      Cannon      &  0.057 &  0.094 \\
      McClure     &  0.075 &  0.057   \\
      Caputo      &  0.150 &  0.046   \\
      Dinescu     &  0.138 &  0.124 \\
      \\
\multicolumn{3}{c}{Median offsets (Adopted minus other)} \\

      Sandage pg\ \ \ \ \ \ \ \  & --0.042 & --0.027   \\
      Sharov      & --0.036 & --0.030   \\
      Cannon      & --0.041 & --0.013 \\
      McClure     & +0.009 & +0.031   \\
      Caputo      & +0.008 & +0.037   \\
      Dinescu     & --0.020 & --0.007 \\
      \\
\multicolumn{3}{c}{$\sqrt{\pi\over2}\,\times\,$mean absolute deviations} \\
     
      Sandage pg\ \ \ \ \ \ \ \  &  0.069 &  0.054   \\
      Sharov      &  0.044 &  0.084   \\
      Cannon      &  0.052 &  0.045 \\
      McClure     &  0.056 &  0.043   \\
      Caputo      &  0.100 &  0.035   \\
      Dinescu     &  0.112 &  0.092 \\
      \\
\multicolumn{3}{c}{Number of stars used in the comparison} \\

      Sandage pg\ \ \ \ \ \ \ \  &    543 &    544   \\
      Sharov      &    74  &    74   \\
      Cannon      &    77  &    77 \\
      McClure     &    585 &    585   \\
      Caputo      &    277 &    277   \\
      Dinescu     &    245 &    245 \\
 \\
\multicolumn{3}{c}{Additive corrections to be applied} \\

      Sandage pg\ \ \ \ \ \ \ \  & --0.040 & --0.025 \\
      Sharov      & --0.038 & --0.028 \\
      Cannon      & --0.037 & --0.010 \\
      McClure     & +0.006 & +0.032 \\
      Caputo      & --0.008 & +0.036 \\
      Dinescu     & --0.018 & --0.006 \\
\enddata
\end{deluxetable}

\clearpage

\begin{deluxetable}{lrrccrrc}
\tablecaption{Estimated precision of photometric studies of \ngc{188}}
\tablecolumns{8}
\tablewidth{0pt}
\tablehead{
\colhead{(1)} & \colhead{(2)} & \colhead{(3)} & \colhead{(4)} & \colhead{(5)} & 
\colhead{(6)} & \colhead{(7)} & \colhead{(8)} \\
\colhead{Sample} & 
\colhead{$\sigma_B$} & 
\colhead{$\sigma_V$} &
\colhead{$\sqrt{\pi\over2}$(m.a.d.)${}_B$} & 
\colhead{$\sqrt{\pi\over2}$(m.a.d.)${}_V$} & 
\colhead{$\bar\sigma$} &
\colhead{$\bar n$} & 
\colhead{$\sigma$(1 obs.)}
}
\startdata

Sandage pe  & 0.047 & 0.048 & 0.033 & 0.032 & 0.040 & 1.8 & 0.060 \\
Eggen ``E'' & 0.035 & 0.029 & 0.035 & 0.029 & 0.032 & 1.6 & 0.045 \\
Eggen ``S'' & 0.045 & 0.034 & 0.038 & 0.029 & 0.036 & 1.4 & 0.048 \\
 \\
Sandage pg  & 0.107 & 0.066 & 0.067 & 0.050 & 0.072 & 1.0 & 0.072 \\
Sharov      & 0.042 & 0.074 & 0.040 & 0.066 & 0.056 & 2.7 & 0.091 \\
Cannon      & 0.057 & 0.094 & 0.052 & 0.045 & 0.062 & 1.0 & 0.062 \\
McClure     & 0.078 & 0.048 & 0.058 & 0.041 & 0.056 & 1.0 & 0.056 \\
Caputo      & 0.089 & 0.030 & 0.060 & 0.027 & 0.058 & 1.1 & 0.054 \\
Dinescu     & 0.120 & 0.110 & 0.104 & 0.087 & 0.105 & 2.0 & 0.149 \\

\enddata
\end{deluxetable}

\clearpage

\begin{deluxetable}{cccccc}
\tablecaption{Fiducial-sequence normal points for \ngc{188}}
\tablecolumns{6}
\tablewidth{0pt}
\tablehead{
\colhead{(1)} & \colhead{(2)} & \colhead{(3)} & \colhead{(4)} & \colhead{(5)} & 
\colhead{(6)} \\
\colhead{$V$} &
\colhead{\bmi} &
\colhead{\umb} &
\colhead{\bmv} &
\colhead{\vmr} &
\colhead{\vmi}
}
\startdata

12.500 & 2.484 & \nodata & 1.241 & 0.676 & 1.243\\
12.647 & 2.450 & \nodata & 1.222 & 0.667 & 1.228\\
12.750 & 2.424 & \nodata & 1.201 & 0.661 & 1.223\\
12.859 & 2.400 & 1.122 & 1.190 & 0.654 & 1.210\\
13.000 & 2.376 & 1.089 & 1.175 & 0.649 & 1.201\\
13.104 & 2.350 & 1.069 & 1.161 & 0.640 & 1.189\\
13.250 & 2.319 & 1.049 & 1.146 & 0.633 & 1.173\\
13.351 & 2.300 & 1.025 & 1.129 & 0.628 & 1.171\\
13.500 & 2.272 & 1.000 & 1.119 & 0.620 & 1.153\\
13.611 & 2.250 & 0.972 & 1.103 & 0.617 & 1.147\\
13.750 & 2.224 & 0.951 & 1.093 & 0.608 & 1.131\\
13.905 & 2.200 & 0.921 & 1.077 & 0.606 & 1.123\\
14.000 & 2.186 & 0.906 & 1.070 & 0.598 & 1.124\\
14.250 & 2.152 & 0.874 & 1.049 & 0.596 & 1.103\\
14.267 & 2.150 & 0.865 & 1.044 & 0.592 & 1.106\\
14.500 & 2.121 & 0.832 & 1.030 & 0.580 & 1.091\\
14.664 & 2.100 & 0.805 & 1.015 & 0.579 & 1.085\\
14.750 & 2.096 & 0.794 & 1.008 & 0.575 & 1.088\\
15.000 & 2.057 & 0.752 & 0.993 & 0.568 & 1.064\\
15.016 & 2.050 & 0.742 & 0.984 & 0.567 & 1.066\\
15.091 & 2.000 & 0.688 & 0.958 & 0.559 & 1.042\\
15.091 & 1.950 & 0.636 & 0.932 & 0.548 & 1.018\\
15.077 & 1.900 & 0.588 & 0.907 & 0.536 & 0.993\\
15.063 & 1.850 & 0.540 & 0.881 & 0.522 & 0.969\\
15.042 & 1.800 & 0.496 & 0.854 & 0.506 & 0.946\\
15.012 & 1.750 & 0.450 & 0.828 & 0.490 & 0.922\\
15.000 & 1.718 & 0.425 & 0.815 & 0.487 & 0.903\\
14.982 & 1.700 & 0.408 & 0.801 & 0.475 & 0.899\\
14.952 & 1.650 & 0.363 & 0.775 & 0.459 & 0.875\\
14.892 & 1.600 & 0.320 & 0.748 & 0.445 & 0.852\\
14.895 & 1.550 & 0.275 & 0.721 & 0.432 & 0.829\\
14.906 & 1.500 & 0.233 & 0.694 & 0.418 & 0.806\\
15.000 & 1.454 & 0.200 & 0.670 & 0.402 & 0.784\\
15.012 & 1.450 & 0.191 & 0.669 & 0.406 & 0.781\\
15.250 & 1.422 & 0.165 & 0.660 & 0.388 & 0.762\\
15.500 & 1.433 & 0.160 & 0.661 & 0.394 & 0.772\\
15.689 & 1.450 & 0.170 & 0.671 & 0.397 & 0.779\\
15.750 & 1.461 & 0.180 & 0.676 & 0.403 & 0.785\\
15.975 & 1.500 & 0.212 & 0.699 & 0.418 & 0.801\\
16.000 & 1.506 & 0.221 & 0.705 & 0.417 & 0.801\\
16.176 & 1.550 & 0.254 & 0.727 & 0.434 & 0.823\\
16.250 & 1.567 & 0.272 & 0.733 & 0.436 & 0.834\\
16.388 & 1.600 & 0.299 & 0.752 & 0.450 & 0.848\\
16.500 & 1.632 & 0.334 & 0.775 & 0.457 & 0.857\\
16.561 & 1.650 & 0.347 & 0.781 & 0.465 & 0.869\\
16.731 & 1.700 & 0.402 & 0.802 & 0.472 & 0.898\\
16.750 & 1.708 & 0.412 & 0.808 & 0.475 & 0.900\\
16.879 & 1.750 & 0.439 & 0.826 & 0.494 & 0.924\\
17.000 & 1.798 & 0.480 & 0.856 & 0.504 & 0.942\\
17.009 & 1.800 & 0.487 & 0.860 & 0.509 & 0.940\\
17.151 & 1.850 & 0.536 & 0.883 & 0.522 & 0.967\\
17.250 & 1.895 & 0.578 & 0.905 & 0.531 & 0.990\\
17.271 & 1.900 & 0.582 & 0.906 & 0.541 & 0.994\\
17.394 & 1.950 & 0.627 & 0.929 & 0.553 & 1.021\\
17.500 & 1.995 & \nodata & 0.951 & 0.563 & 1.044 \\
17.518 & 2.000 & 0.674 & 0.956 & 0.567 & 1.044\\
17.641 & 2.050 & 0.725 & 0.979 & 0.585 & 1.071\\
17.750 & 2.102 & \nodata & 1.011 & 0.606 & 1.091\\
17.751 & 2.100 & 0.773 & 1.009 & 0.608 & 1.091\\
17.867 & 2.150 & \nodata & 1.028 & 0.630 & 1.122\\
17.976 & 2.210 & \nodata & 1.050 & 0.642 & 1.160\\
18.000 & 2.224 & \nodata & 1.056 & 0.651 & 1.168\\
18.082 & 2.265 & \nodata & 1.070 & 0.656 & 1.195\\
18.138 & 2.300 & \nodata & 1.080 & 0.671 & 1.220\\
18.250 & 2.350 & \nodata & 1.100 & 0.694 & 1.247\\
18.310 & 2.400 & \nodata & 1.124 & 0.711 & 1.276\\
18.453 & 2.465 & \nodata & 1.157 & 0.732 & 1.308\\
18.500 & 2.500 & \nodata & 1.173 & 0.748 & 1.327\\
18.622 & 2.550 & \nodata & 1.191 & 0.762 & 1.359\\
18.714 & 2.600 & \nodata & 1.214 & 0.776 & 1.386\\
18.750 & 2.621 & \nodata & 1.224 & 0.786 & 1.397\\
18.784 & 2.650 & \nodata & 1.235 & 0.793 & 1.415\\
18.887 & 2.700 & \nodata & 1.251 & 0.810 & 1.449\\
18.952 & 2.750 & \nodata & 1.272 & 0.822 & 1.478\\
19.000 & 2.780 & \nodata & 1.291 & 0.827 & 1.489\\
19.056 & 2.800 & \nodata & 1.299 & 0.841 & 1.501\\
19.141 & 2.850 & \nodata & 1.311 & 0.851 & 1.539\\
19.233 & 2.900 & \nodata & 1.329 & 0.862 & 1.571\\
19.250 & 2.907 & \nodata & 1.337 & 0.867 & 1.570\\
19.325 & 2.950 & \nodata & 1.355 & 0.872 & 1.595\\
19.420 & 3.000 & \nodata & 1.369 & 0.884 & 1.631\\
19.500 & 3.042 & \nodata & 1.384 & 0.895 & 1.658\\
19.500 & 3.050 & \nodata & 1.393 & 0.901 & 1.657\\
19.600 & 3.100 & \nodata & 1.409 & 0.906 & 1.691\\
19.704 & 3.150 & \nodata & 1.429 & 0.917 & 1.721\\
19.750 & 3.177 & \nodata & 1.439 & 0.926 & 1.742\\
19.772 & 3.200 & \nodata & 1.445 & 0.929 & 1.755\\
19.860 & 3.240 & \nodata & 1.454 & 0.940 & 1.786\\
20.000 & 3.296 & \nodata & 1.468 & 0.950 & 1.828\\
20.020 & 3.300 & \nodata & 1.470 & 0.954 & 1.830\\
20.129 & 3.350 & \nodata & 1.489 & 0.963 & 1.861\\
20.235 & 3.400 & \nodata & 1.506 & 0.974 & 1.894\\
20.250 & 3.428 & \nodata & 1.509 & 0.979 & 1.919\\
20.313 & 3.450 & \nodata & 1.525 & 0.983 & 1.925\\
20.397 & 3.500 & \nodata & 1.539 & 0.994 & 1.961\\
20.474 & 3.550 & \nodata & 1.554 & 1.007 & 1.996\\
20.500 & 3.577 & \nodata & 1.559 & 1.009 & 2.018\\
20.547 & 3.600 & \nodata & 1.564 & 1.013 & 2.036\\

\enddata
\end{deluxetable}

% \clearpage
% \plotone{stetsonfig1.eps}

% \clearpage
% \plotone{stetsonfig2.eps}

% \clearpage
% \plotone{stetsonfig3.eps}

% \clearpage
% \plotone{stetsonfig4.eps}

% \clearpage
% \plotone{stetsonfig5.eps}

% \clearpage
% \plotone{stetsonfig6.eps}

% \clearpage
% \plotone{stetsonfig7.eps}

% \clearpage
% \plotone{stetsonfig8.eps}

% \clearpage
% \plotone{stetsonfig9.eps}

% \clearpage
% \plotone{stetsonfig10.eps}

% \clearpage
% \plotone{stetsonfig11.eps}

\end{document}